# Mapping the directional emission of quasi-2D photonic crystals of semiconductor nanowires using Fourier microscopy

Yannik Fontana (1,5), Grzegorz Grzela (1), Erik P. A. M. Bakkers (2,3,4), Jaime Gómez Rivas (1,2)

(1) FOM Institute AMOLF, c/o Philips Research, High-Tech Campus 4, 5656 AE Eindhoven, (The Netherlands)

(2) COBRA Research Institute, Eindhoven University of Technology, P.O. Box 513, 5600 MB Eindhoven (The Netherlands)

(3) Kavli Institute of Nanoscience, Quantum Transport, Delft University of Technology, 2600 GA Delft, (The Netherlands)

(4) Philips Research Laboratories, High-Tech Campus 4, 5656 AE Eindhoven (The Netherlands)

(5) Current affiliation: Laboratoire des Matériaux Semiconducteurs, Institut des Matériaux, Ecole Polytechnique Fédérale de Lausanne, 1015 Lausanne, (Switzerland)

## Abstract

Controlling the dispersion and directionality of the emission of nanosources is one of the major goals of nanophotonics research. This control will allow the development of highly efficient nanosources even at the single photon level. One of the ways to achieve this goal is to couple the emission to Bloch modes of periodic structures. Here, we present the first measurements of the directional emission from nanowire photonic crystals by using Fourier microscopy. With this technique we efficiently collect and resolve the directional emission of nanowires within the numerical aperture of a microscope objective. The light emission from a heterostructure grown in each nanowire is governed by the photonic (Bloch) modes of the photonic crystal. We also demonstrate that the directionality of the emission can be easily controlled by infiltrating the photonic crystal with a high refractive index liquid. This work opens new possibilities for the control of the emission of sources in nanowires.

# I. Introduction

The control of the light emission of nanostructures in terms of efficiency and directionality is one of the major paradigms of nanophotonics. This control will facilitate the design of sources for applications ranging from LEDs, integrated photonics and single photon sources for quantum information. Among the most promising structures for light emission are photonic crystals (PhCs), i.e., periodic structures of two or more dielectrics or semiconductors with different refractive indices.[1] The lattice constant of such a structure is on the order of the wavelength of light. PhCs have become a broad field of research over the past years. The periodic refractive index acts as a periodic potential for electromagnetic waves. The propagation of electromagnetic waves in such a medium is described by Bloch modes. Typically, the dispersion relations of Bloch modes in PhCs reveal photonic stop-gaps, i.e., frequency ranges in which there are no Bloch modes available for light propagation along certain directions. A photonic band-gap is formed when propagation is forbidden along all directions. In a perfect PhC with a band-gap, the spontaneous emission of an excited state can be, in theory, fully inhibited. These properties have been extensively studied to achieve control over the spontaneous light emission of nanophotonic sources.[2]

PhCs have been proven to confine and manipulate light,[3, 4] modify the density of optical states,[5, 6] create high quality factor cavities leading to lasing,[7] and increase the outcoupling efficiency of light sources.[8-10]

Semiconductor nanowires are potentially interesting building blocks for two dimensional (2D) PhCs. Nanowires grown by the vapor-solid-liquid (VLS) method have become a major field of research due to their unique optical properties.[11-15] The bottom-up approaches which are used to grow nanowires allow the combination of different semiconductors in single nanostructures despite the mismatches of the crystal lattice constant. This property has opened new opportunities to design a large variety of heterostructures. Some examples of these structures are wrapped-around quantum wells,[16, 17] superlattices,[18, 19] quantum dots embedded in nanowires showing single photon emission,[20-23] and zincblende-wurtzite phase homostructures.[24, 25] Furthermore, it is possible to grow III-V semiconductor nanowires on top of silicon substrates, combining the high-end properties of III-V material with well-developed and cheap silicon technology.[26] It has been recently demonstrated that ordered arrays of nanowires can be grown on patterned substrates.[27-29] These ordered arrays of nanowires behave as 2D PhCs, modifying the nanowire emission,[30] and even leading to lasing.[31]

In this manuscript, we demonstrate the directional emission of heterostructures embedded in semiconductor nanowires forming a quasi-2D PhC using Fourier microscopy. This technique allows us to image the emission in the reciprocal space in which each coordinate corresponds to one direction of emission. The directional emission arises from the coupling of the emission from heterostructures in individual nanowires to Bloch modes in the PhC.[8, 32] The investigated samples can be described as a quasi-2D PhC due to the periodic arrangement of the nanowires and their finite length.[1] This finite length allows an efficient coupling of the photoluminescence of heterostructures into free space at certain directions. We also demonstrate that it is possible to efficiently control the directionality of the emission by changing the refractive index of the medium surrounding the nanowires. This change is achieved by infiltration of the nanowire array with a liquid. The introduction of Fourier microscopy for the study of the emission of nanowires will facilitate the design of efficient and directional light sources that could be extended to single photon emission and electroluminescence.

## II. Sample description

The nanowire arrays are grown on top of a thick (111)B InP substrate by gold-catalyzed vapor-liquid-solid (VLS) mechanism in a MOVPE reactor.[33] The exact position of the nanowires is defined by the position of the gold particles catalyzing the growth. These particles were deposited by using substrate conformal imprint lithography,[34] followed by a lift-off process. The complete fabrication process of this sample is discussed elsewhere.[29] The array used in this work is formed by InP nanowires embedding an $InAs_{0.1}P_{0.9}$ heterostructure. The incorporation of the As in InP narrows the semiconductor's electronic band gap and red-shifts the photoluminescence. The nanowires have an overall length of 3 μm, with a tapered base (1 μm long) and straight upper part (2 μm long). The InAsP heterostructure, with a length of 20 nm, is located at 1 μm from the top of the wires. The diameters at the bottom of the base and of the straight part are 270 nm and 90 nm, respectively. The nanowires are arranged in a square lattice with a pitch of 513 nm. This structure can be appreciated in the tilted scanning electron microscope image of the sample shown in Fig. 1(a). Defects in the periodic structure consist of slightly tilted and missing nanowires. These defects do not have a significant influence on the optical measurements presented below and they are not considered further. Defects in the crystallographic structure of individual nanowires may influence the intensity of the emission and broaden the emitted spectrum, which is irrelevant for the directional emission of the periodic lattice of nanowires.

The photoluminescence (PL) spectrum of the nanowire array is shown in Fig. 1(b). The peak centered at $\lambda$ = 870 nm can be attributed to the PL from the InP in the wurtzite phase. InP nanowires grow preferentially in wurtzite phase in contrast to bulk zincblende InP (as in the case of the substrate) that is expected to emit at around $\lambda$ = 922 nm due to the lower band gap energy.[35] The broader, red-shifted peak centered at $\lambda$ = 970 nm, corresponds to the PL from the 20 nm thick InAsP heterostructure incorporated in the nanowires. This heterostructure provides a light source embedded in the periodic array and with a well-defined position. The large size of the InAsP heterostructure prevents any noticeable quantum confinement for the electrons and its electronic band structure, as well as the emission properties, is assumed to be that of bulk InAsP.

## III. Experimental details

A 100x/0.95 PL APO (Leica Microsystems) and a 100x/1.3 HCX PL Fluotar (Leica Microsystems) objectives were used for the measurements in respectively air and immersion oil (n = 1.518). The array was optically pumped through the same objective used to collect the PL. We have used a continuous wave diode-pumped solid-state laser with an emission wavelength centered at 785 nm to excite the nanowire array. This wavelength allows us to efficiently pump the heterostructure, avoiding a strong decrease of the pump intensity by absorption in the top section of the InP nanowires.[36] The excitation used in the measurements shown in section V is linearly polarized along the square lattice axis. In order to rule out any influence of the pump polarization on the emission pattern, several measurements with the plane of polarization oriented in 3 different directions every 45° were performed. These measurements (not shown here) did not reveal any differences between the emission patterns. The only differences were on the emission intensities.

The directional emission of the nanowire array was mapped using the Fourier transform properties of lenses. Light emitted at a defined angle and coupled into an objective will be focused onto a unique point at its back focal plane (Fourier plane). This concept is schematically illustrated in Fig. 2(a), where the emission of an extended source at different angles is collected by an objective lens and imaged onto the Fourier plane. Hence, by imaging the Fourier plane of the objective it is possible to determine the emission pattern of the emitter. The emission into the solid angle collected by the microscope objective appears as a disc in the Fourier image. Each point in this disc corresponds to one direction of emission, e.g., a spot located at the edge of the Fourier image represents the emission at a large angle, given by the numerical aperture (NA) of the objective, relative to the normal to the sample surface. The center of the image corresponds to the emission normal to the sample surface. One remarkable characteristic of Fourier imaging, as compared to real-space imaging, is the shift invariance property: The spatial coordinates on the Fourier plane depend only on the associated set of azimuthal and elevation angles of the emission and are independent of the distance between the source and the Fourier lens. For this reason, even if the emitter is placed out of the focal plane of the Fourier lens, its emission pattern in the Fourier image will remain unchanged within the shift invariance range of the measurement apparatus.[37]

In practice, the imaging of the back focal plane of the objective is done with the help of a lens of focal length f placed at a distance 2f from the back focal plane and 2f from a CCD camera. With this configuration a 1:1 image of the Fourier plane is projected onto the CCD. In order to maximize the angular range of the measurements, the NA of the objective should be as large as possible. For our measurements of the nanowire PhC in air, the maximum collection angle associated with a 100x objective with a NA of 0.95 is 72°. The measurements of the infiltrated PhC were done with a 100x immersion oil objective with a NA of 1.3. This objective is used with immersion oil of refractive index of 1.52, which leads to a maximum collection angle of 59°. Our Fourier microscope is also equipped with a polarizer in front of the CCD camera, i.e., in the detection path. This polarizer makes it possible to distinguish between s- and p-polarized emission along the azimuthal line parallel (p-polarized) and perpendicular (s-polarized) to the transmission axis of the polarizer. The transmission through the polarizer in other directions is a superposition of p- and s-polarized emission (see Fig. 2(b)).

The Fourier images were recorded on a CCD camera. Integration times were of the order of 50 ms for the InP emission and 1 s for the InAsP emission. In order to distinguish the PL emission from the wurtzite InP nanowires from the PL emission of the InAsP heterostructures embedded into the nanowires, the Fourier images were measured after transmission of the PL through a band pass filter. The nanowire emission was measured after a filter with a central wavelength of 880 nm and a bandwidth of 10 nm at FWHM, while the emission from the heterostructures was measured after a filter with a central wavelength of 960 nm and a bandwidth of 10 nm at FWHM. The transmission of these filters is illustrated in the spectrum of Fig. 1(b), where the cyan dashed area corresponds to the sample PL after the 880 nm filter and the red dashed area represents the PL after the 960 nm filter. The choice of the filters was determined by the emission spectra of InP and InAsP and by their availability.

The images for unpolarized light are the sum of the images taken with 2 orthogonal polarizations. In the unpolarized images, the intensity along the horizontal direction appears lower. This effect is not the feature of the directional emission, but it is attributed to the polarization dependent transmission through a beam splitter used in the setup.

For the measurements described in section V.B. the infiltration of the nanowire PhC was done as follows: Immersion oil was pipetted on top of the array in order to cover the whole sample. To force the infiltration of the voids in between nanowires, the sample was placed into a vacuum tube during 5 minutes under a pressure on the order of 0.02 atm. The optical measurements were performed after this procedure and within 5 hours. A second set of measurements was realized 24 hours later and no significant changes were noticed. The PL of both InP and InAsP was not quenched by the contact with the immersion oil.

## IV. Photonic band structure calculations

Light propagation in a structure formed by different dielectrics is dictated by multiple scattering and interference. If the structure is periodic, the scattering and interference lead to Bloch modes, which represent the eigenstates of the PhC. The eigenvalues associated with the Bloch modes define the PhC band structure or the dispersion relation of the light propagating in the periodic structure. Figure 3(a) displays a calculation of the band structure of a 2D PhC formed by infinitely long InP cylinders in air ($n_{InP}$ = 3.37, $n_{Air}$ = 1). The eigenmodes of Maxwell's equations in the periodic dielectric structure were computed in the frequency domain by the plane wave expansion method, using the freely available software package MIT Photonic Bands.[38] The band structure is represented as a function of the wavenumber parallel to the plane containing the reciprocal lattice vectors of the 2D crystal, $k_{||}$, where the $\Gamma$ point corresponds to $k_{||}$ = 0 and the X and M points are the edges of the irreducible Brillouin zone. The different bands in Fig. 3(a) have been labeled depending on the polarization of the electromagnetic wave in the structure. Transverse Electric (TE) Bloch modes correspond to a polarization perpendicular to nanowire elongation ($E_z$ = 0), while Transverse Magnetic (TM) modes have a polarization parallel to the nanowires ($E_x = E_y = 0$). The modes are calculated for light propagating in plane perpendicular to the nanowires. Light coupled to a mode can be scattered out of the PhC conserving the parallel wavenumber at the interface (as shown in Fig. 3(b)). In this way, we can associate the isofrequency surfaces, which are planes of constant frequency in the band structure, with the direction of light emission out of the array. Figures 3(c) and (d) represent the isofrequency surfaces for $\lambda$ = 880 nm and 960 nm respectively. The left panel in these figures corresponds to TM modes, while the right panel contains TE modes. We have also indicated in this figure the order of the mode. We must point out that our periodic array of InP nanowires is not truly a 2D-PhC due to the finite length of the nanowires. Nevertheless and as it is shown later, the directional emission of the photoluminescence from the sample is determined by its periodicity and can be associated with the band structure of the infinite crystal. Therefore, we refer to the array as a quasi-2D photonic crystal.

## V. Results and discussion

### A. Nanowire array in air

The Fourier image of the InP nanowire emission at $\lambda$ = 880 nm without polarization selection is displayed in Fig. 4(a). This emission pattern strongly differs from a Lambertian emission expected from thin emitting films and flat surfaces. The four-fold rotational symmetry of the Fourier image follows the symmetry of the square array of nanowires. The most salient feature is the emission bands at large angles. In order to determine the polarization of the emission in these bands, we have obtained the Fourier image with the polarizer oriented along the vertical direction (azimuthal angle φ = 90° and 270°) as it is displayed in Fig. 4(b). The emission along φ = 90° and 270° corresponds to p-polarized emission, while the emission along φ = 0° and 180° is s-polarized. The absence of the emission bands for s-polarization allows concluding that these bands can be attributed to the second order TM band of the PhC (see Fig. 3(c)).

As can be appreciated in Fig. 4(b), each emission band is formed by two concentric curves. These curves are marked by the arrows in the cut to the Fourier image at φ = 90° and 270° displayed on the right inset of the figure. The origin of the double bands might be attributed to the finite length of the nanowires. The emission at $\lambda$ = 880 nm comes from the whole InP nanowire, including the uppermost section, close to the sample interface, where the optical pump is absorbed.[36] The sample interface plays an important role in the light emission by the InP nanowires: An interface separating a photonic crystal from the homogenous medium where the emission is detected, is present in any experiment. The conservation of the parallel wavenumber to the interface makes it possible to correlate the angular emission with the isofrequency surfaces of the Bloch modes of the PhC. However, the interface breaks the translational invariance of the nanowire array along their elongation and limits the validity of the band structure calculation done for an infinite 2D-PhC. Therefore, it is expected that light emitted by the nanowire layer close to the interface will be affected by the presence of this interface. For angles θ up to ≈35°, there is a relatively strong emission with no particular directional features. We associate this to the emission of the nanowires that is not coupled to the modes of the periodic structure.

In order to minimize the influence of the interface in the emission of the nanowire PhC, we consider in Fig. 4 (c) and (d) the unpolarized and polarized emission from the InAsP heterostructure, located at a depth of 1 μm from the surface. In these measurements we observe a unique set of bands, corresponding to an emission around θ ≈ 57° for φ = 0°, 90°, 180° and 270°. The cut in Fig. 4(d) at φ = 90° and 270° (right inset) reveals a minimum in the emission at θ ≈ 50°, before reaching the maximum at θ ≈ 57°. Note that there is a small asymmetry in the maximum intensity emitted at θ ≈ 57° for φ = 90° and 270°. This asymmetry can be attributed to non-uniform collection efficiency in our setup. In Ref. [30], the directional emission of similar arrays of nanowires has been reported and described with finite difference time domain (FDTD) simulations.[30] The directional emission was explained by its coupling to modes in the periodic structure and Fabry-Pérot resonances in the nanowire layer with finite length. The emission in Figs. 4(b) and (d) matches well with the isofrequency surfaces of the $TM_2$ mode at λ = 960 nm in a 2D photonic crystal (Fig. 3(d)) with the same pitch and dielectric rod diameter as the nanowire array. The absence of the $TE_2$ mode emission in the Fourier images can be explained by looking at the spatial distribution of the electric field intensity in the unit cell of the photonic crystal. Figure 5 illustrates a calculation of this field intensity distribution, using the plane wave expansion method,[38] where (a) and (b) corresponds to the $TM_2$ mode and (c) and (d) to the $TE_2$ mode at the two investigated wavelengths, λ = 880 nm ((a) and (c)) and λ = 960 nm ((b) and (d)). TM modes are described by the $E_z$ component of the field, while $E_y$ describes the TE modes in Γ-X direction of the reciprocal space (see the inset of Fig. 3(a)). At λ = 880 nm, most of the field intensity of the $TM_2$ mode is concentrated in the cylinder, in contrast to the $TE_2$ mode at the same wavelength, where the maximum field intensity is located around the cylinder. Similar behavior can be appreciated at λ = 960 nm, though, the contrast between the field intensity distributions for $TM_2$ and for $TE_2$ is lower. Therefore, it is reasonable to expect that light generated by the recombination of photo-excited electron-hole pairs in the nanowire can more easily couple to the $TM_2$ mode due to the good overlap of this mode with the nanowire.

## B. Nanowire array infiltrated with a high-refractive-index liquid

The band structure of a PhC depends on the geometrical parameters defining the periodic structure, as well as on the permittivity of its constituents. Therefore, the emission pattern is expected to change when the refractive index of the medium surrounding the nanowires is varied. Figures 6 (a) and (b) illustrates this modified emission by plotting the isofrequency surfaces of a 2D PhC of InP nanowire with the same geometry as in Fig. 3 (c) and (d), but with the voids in between nanowires filled with a medium with a refractive index of 1.52. Figure 6 (a) corresponds to the isofrequency surface at λ = 880 nm, while (b) is the isofrequency surfaces at λ = 960 nm. The band structure is much more complex and many TE and TM bands are expected to influence the emission of the array. Figures 7 (a) and (b) show the Fourier images measured for the unpolarized emission from the sample infiltrated with the immersion oil (n = 1.52) at these two wavelengths. These Fourier images show a 4-fold symmetry but are very different from the case of the uninfiltrated array. The emission at λ=880nm from the InP nanowire (Figs. 7(a) and (c)) is dominated by the non-directional emission which is not coupled to the modes of the periodic structure. The emission pattern of the heterostructure at $\lambda$=960nm (Figs. 7(b) and (d)) is much richer in directional features due to the more complicated band structure shown in Fig. 6(b). Clearly, changing the refractive index of an embedding medium modifies the dispersion relation of the quasi-2D photonic crystal influencing the directional light emission from embedded nanosources. While the uninfiltrated photonic crystal emission bands manifest mainly along the vertical and horizontal direction in the Fourier image, the infiltrated structure reveals more bands along the intermediate directions. For this reason we took Fourier images with a polarizer's axis oriented along the azimuthal angle of $\varphi$ = 135° and 315°. For both investigated wavelengths, $\lambda$ = 880 nm and $\lambda$ = 960 nm, the emission bands show polarization dependence (Fig. 7 (c) and (d)). Unlike before, there is a contribution of s-polarized directional emission features in the Fourier images. This contribution might be caused by the coupling of the emission to TE modes of the infiltrated sample or by scattering at the interface of the quasi-2D photonic crystal.

## Conclusions

In conclusion, we have demonstrated the modified emission of semiconductor nanowire arrays and heterostructures in nanowires by scattering in a periodic structure. In particular, we have measured the directional photoluminescence of 2D nanowire array by using Fourier microscopy. This emission has been correlated to Bloch modes in 2D photonic crystals. By filling the voids between the nanowires, it is possible to modify the emission.

The use of Fourier microscopy for the study of nanowires will facilitate the design of directional photon sources in quantum information and integrated optics applications.

## Acknowledgements

We are grateful to M. Hocevar and A. Pierret for providing us with the sample and for useful discussions. We are also grateful to S. L. Diedenhofen, O. T. A. Janssen and I. Seršić for fruitful discussions. This work is part of the research program of the "Stichting voor Fundamenteel Onderzoek der Materie (FOM)", which is financially supported by the "Nederlandse organisatie voor Wetenschappelijk Onderzoek (NWO)" and is part of an industrial partnership program between Philips and FOM.

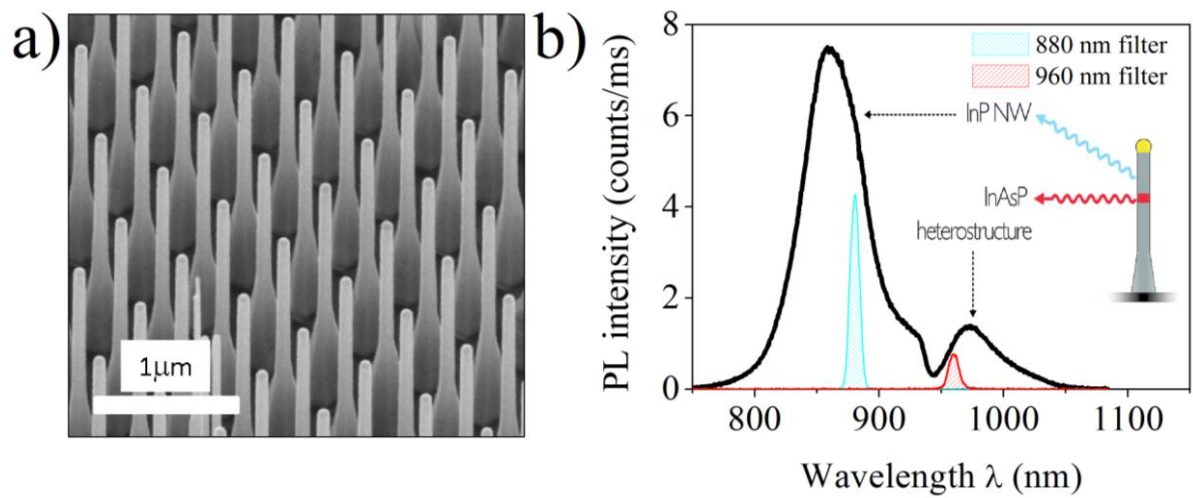

FIG. 1. (a) Scanning electron micrograph of the array of InP nanowires taken at an inclination angle of 30°. Each nanowire in the array has a length of 3μm and contains a 20nm InAsP segment located at 1μm from the top. (b) Photoluminescence spectrum of the InP nanowire array. The broad peak centered at λ = 860 nm is the emission from the wurtzite InP that forms the nanowires. The peak centered around λ = 970 nm is associated with the emission from the InAsP segment located 1μm from the top of each nanowire (as represented in the inset). Directionality of the emission was investigated using 2 different band pass filters: λ = 880 nm and λ = 960 nm, both with a transmission band of 10 nm. The PL spectra of the nanowire array measured with the filters are displayed in the graph with cyan and red dashed areas, respectively.

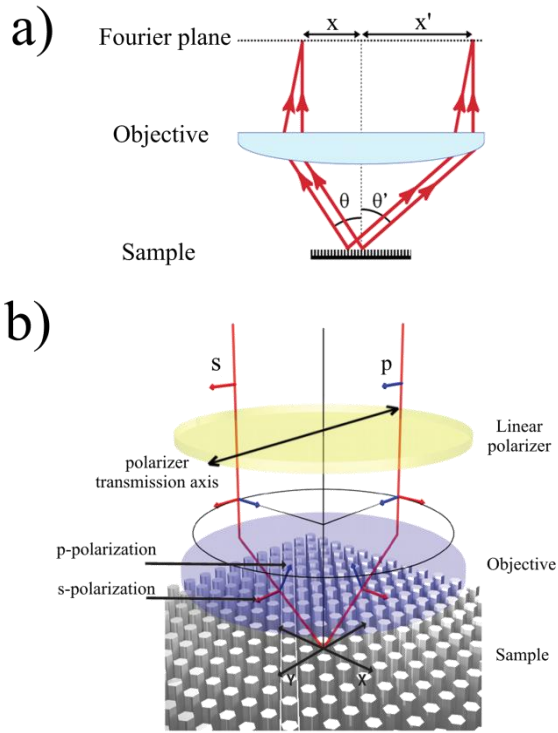

FIG. 2. (a) Working principle of a microscope in Fourier mode. The emission in a certain direction is captured by the objective and focused on an unique point on the back focal plane of the objective (Fourier plane). Each point in the Fourier plane represents a unique set of elevation and azimuthal angles of the emission. (b) Polarization analysis of the Fourier images. In the plane of incidence of the sample, the electric field vector of the p-polarized emission always points toward the center of a Fourier image. Strictly along the polarizer's transmission axis, the transmitted light is purely p-polarized with respect to the sample surface. This means that the purely p-polarized emission is the profile of the Fourier image through its center and along the indicated polarizer's transmission axis. Consistently, only the profile in the direction perpendicular to the polarizer's transmission axis shows purely s-polarized emission. Along every other azimuthal direction, the transmitted light is a superposition of the s and p–polarized emission.

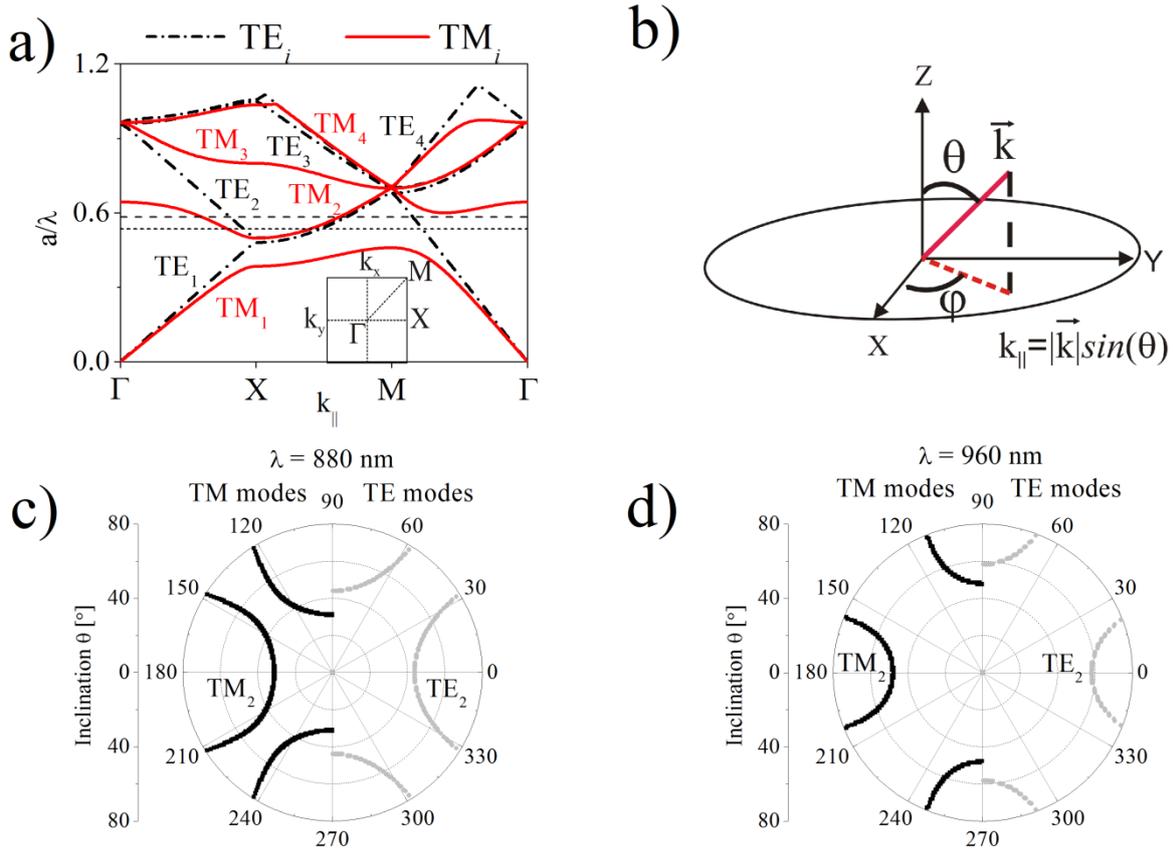

FIG. 3. (a) Band structure of a 2D photonic crystal with the same lattice parameters as in the array showing the TM (solid curves) and TE Bloch (dash-dotted curves) modes between high symmetry points of the first Brillouin zone in the reciprocal space. The dashed and dotted horizontal lines indicate the wavelengths of λ = 880 nm and λ = 960 nm, respectively. The inset illustrates the reciprocal space of the PhC within the first Brillouin zone with the symmetry points. (b) Illustration of polar coordinates used to plot Fourier images. In those images, the radius represents the angle of emission θ, while the azimuthal angle φ corresponds to the azimuthal angle of the emission. The angle θ is related to the magnitude of the wavenumber in plane of the sample $k_∥$, while the angle φ is related to the ratio of the in-plane wavenumber components, namely $k_x$ and $k_y$. (c) Calculated isofrequency surface for λ = 880 nm, the wavenumbers have been transformed into polar coordinates. The TM modes are displayed on the left, while the TE modes are on the right. The isofrequency surface for λ = 960 nm is shown in (d).

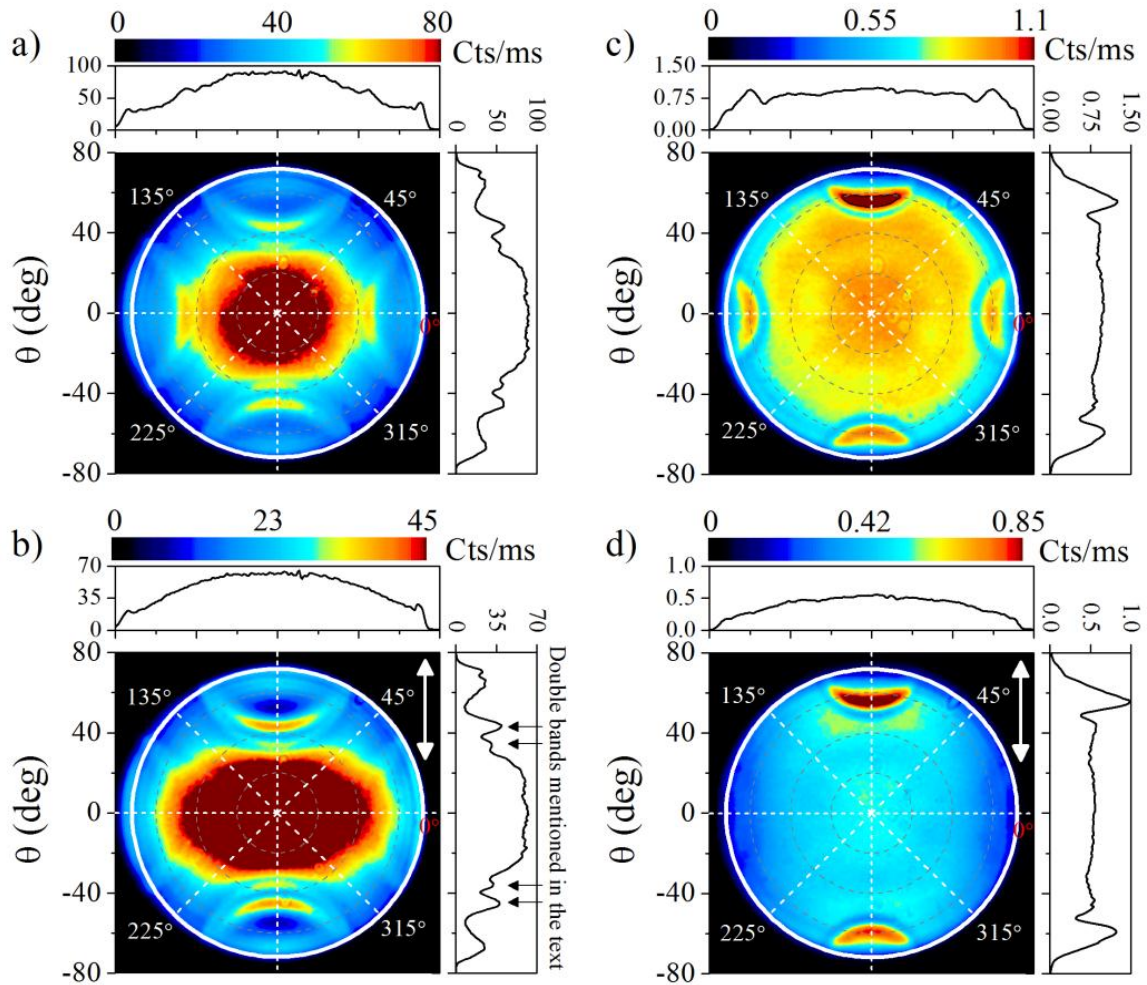

FIG. 4. Measured Fourier images of the light emission from an InP nanowire array emitting in air. The top images represent the unpolarized emission for λ = 880 nm (a) and λ = 960 nm (c) while the bottom ones (b and d) are taken for the same wavelengths but with a polarizer in front of the Fourier camera. The polarizer's transmission axis is aligned along the vertical direction (φ = 90° and 270°) and is indicated in the graphs by the white double-arrows. The white circle represents the maximum collection angle of the objective with NA = 0.95, equal to θ = 72°. The graphs on the top and right side of each figure show the profiles of the Fourier images along the horizontal and vertical directions crossing at the center of the images.

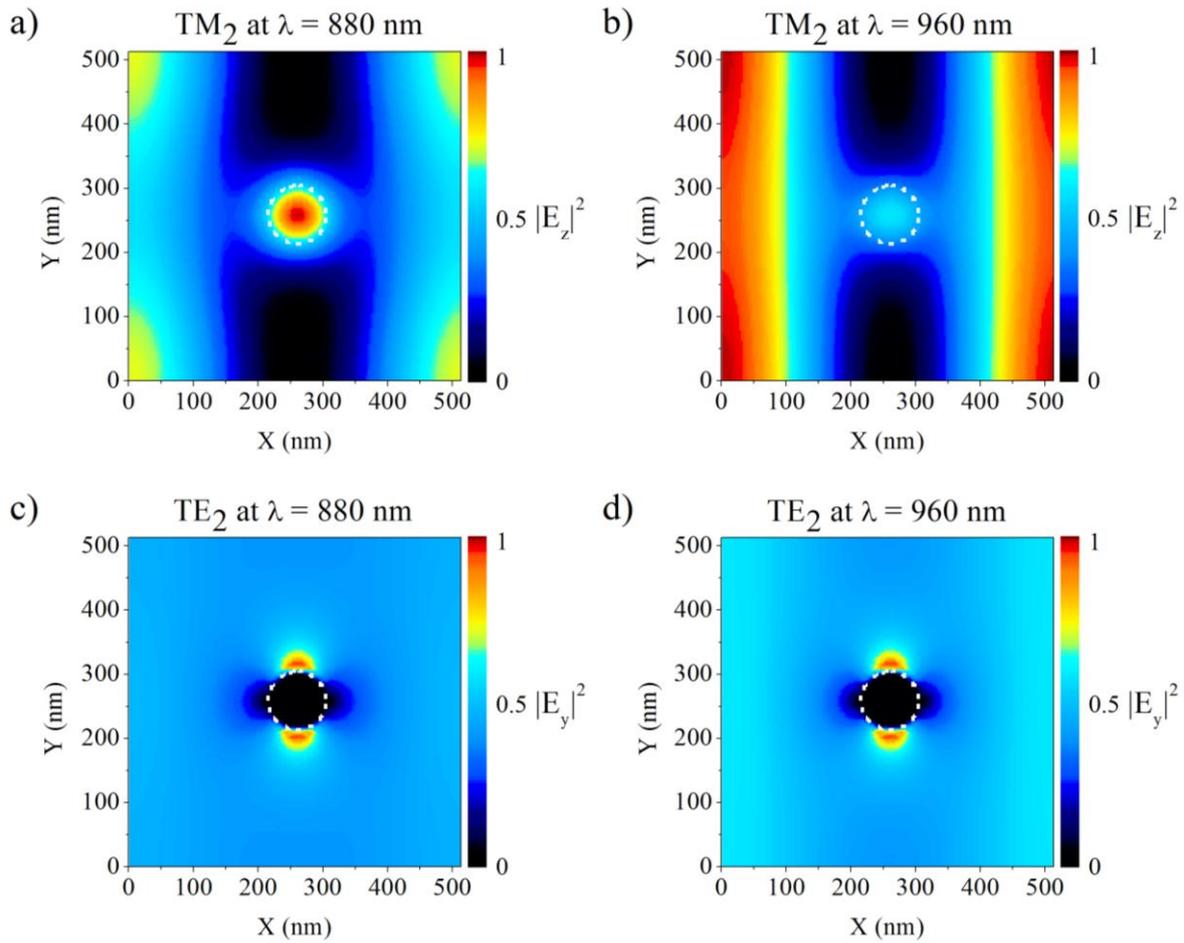

FIG. 5. Calculated electric field intensity distribution of the photonic modes in a unit cell of a 2D photonic crystal. a) $E_z$ field intensity distribution of the $TM_2$ mode at $\lambda$ = 880 nm and b) $\lambda$ = 960 nm. c) $E_y$ field intensity distribution of the $TE_2$ mode at $\lambda$ = 880 nm, and d) at $\lambda$ = 960 nm. In all graphs, the field intensity is normalized to a maximum. The white dashed lines mark the position of the dielectric cylinder with a refractive index of n = 3.37 embedded in vacuum with a refractive index of n = 1. The field intensity distributions are calculated for the mode wave numbers in the Γ-X direction in the reciprocal space.

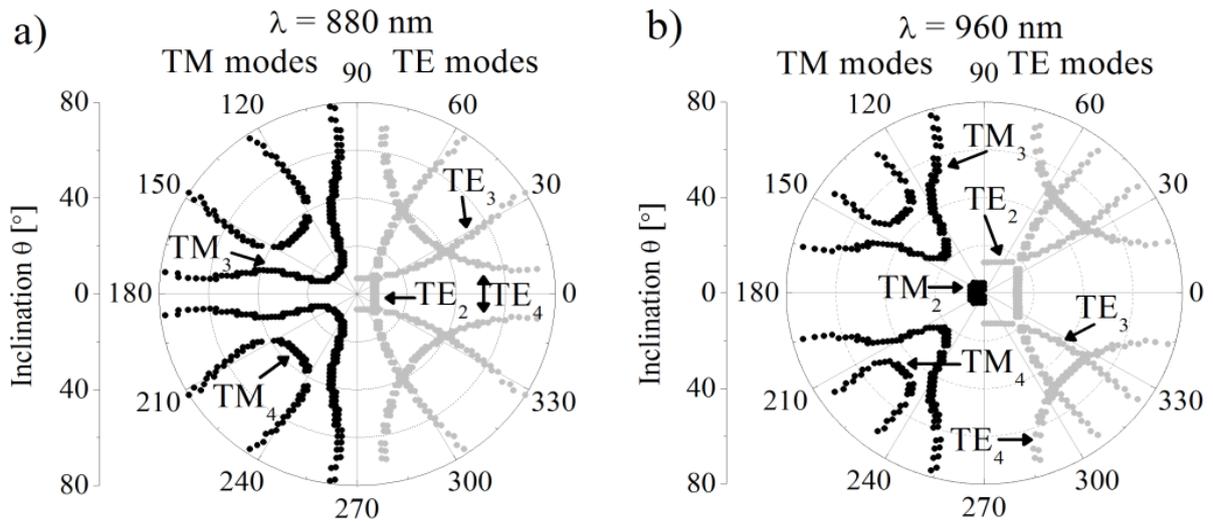

FIG. 6. Calculated isofrequency surfaces of a 2D photonic crystal formed by infinitely long cylinders embedded in a medium with a refractive index of n = 1.52. The cylinders are made of a dielectric material with a refractive index of n = 3.37. (a) represents the isofrequency surface for λ = 880 nm, while (b) shows the isofrequency surface for λ = 960 nm. The TM modes are shown on the left side of the graphs and the TE modes are presented on the right side.

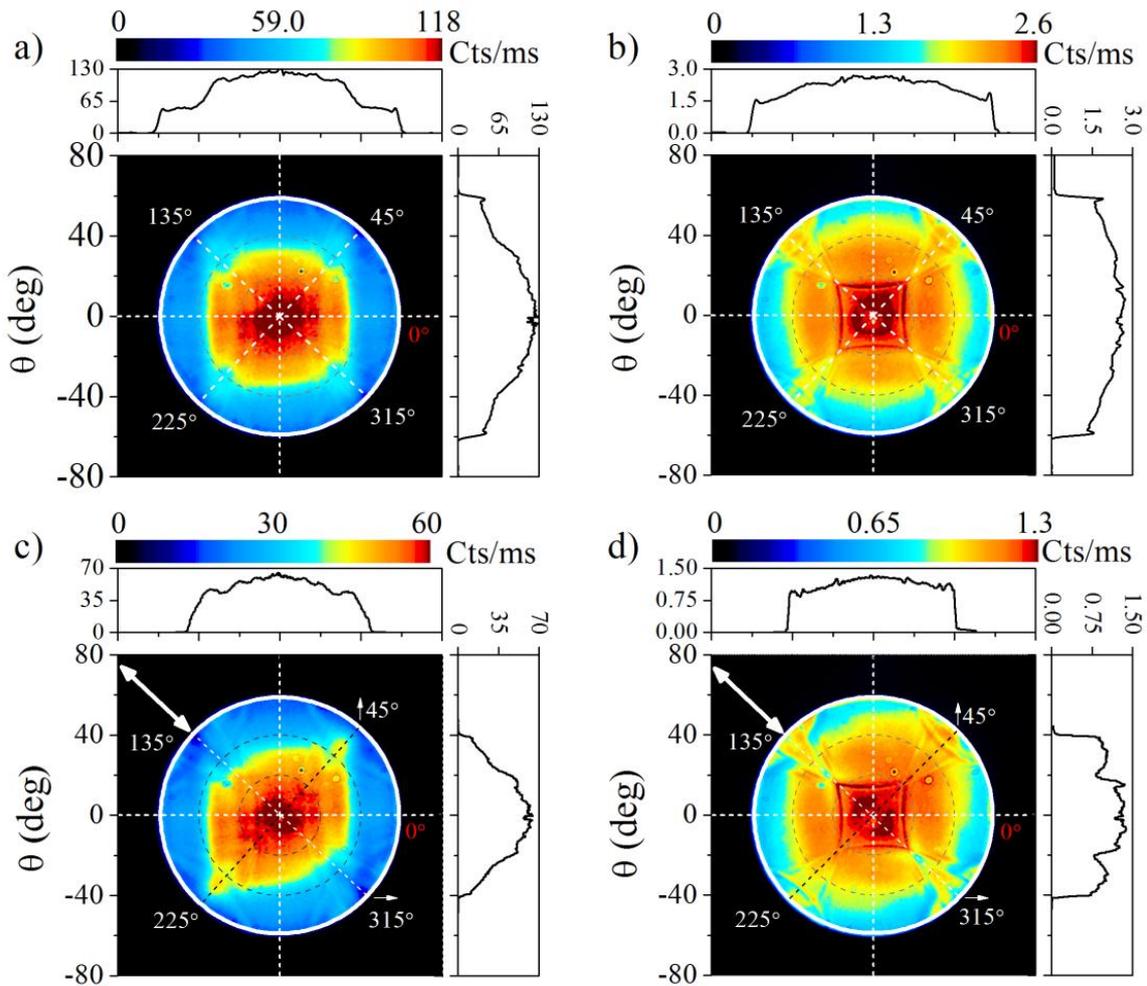

FIG. 7. Measured Fourier images of the photoluminescence emission from the InP nanowire array embedded in an immersion oil with refractive index of n = 1.52. The top graphs (a and b) present the images of the unpolarized emission taken with band pass filters with a central wavelength of λ = 880 nm (a) and λ = 960 nm (b). The bottom graphs (c and d) display the emission for the same wavelengths recorded with a polarizer in front of the Fourier camera. The polarizer's transmission axis is oriented at φ = 135° and 315° and it is indicated by the white double-arrows in both graphs. The white circle represents the maximum collection angle, which for the objective with NA = 1.3 in a medium of n = 1.518 is equal to θ = 59°. The graphs at the top and right side of figures (a) and (b) show the profiles of the Fourier images along the horizontal and vertical directions crossing at the center of the images. In (c) and (d) the profiles of the Fourier images are plotted along φ = 45° and 225° (top graph,

corresponding to the black dashed line) and along φ = 135° and 315° (right graph, corresponding to the white dashed line) to illustrate the differences between pure s- and p-polarized emission. These diagonal profiles are projected on the horizontal and vertical axes, respectively, appearing narrower than the profiles in (a) and (b).